\documentclass[twoside,slac_one]{revtex4}
\usepackage{graphicx}
\usepackage{fancyhdr}
\usepackage{amsmath} % American Mathematics Society standards
\usepackage{bm}% bold math
\usepackage{amsxtra}
\usepackage{amssymb}
\usepackage{amsthm}
\usepackage{latexsym}
\usepackage{lscape}

\newcommand{\be}{\begin{eqnarray}}
\newcommand{\ben}{\begin{eqnarray}\nonumber}
\newcommand{\ee}{\end{eqnarray}}

\begin{document}

%\title{\font\cmr12 String Landscape and Supernovae Ia}
\title{ String Landscape and Supernovae Ia}


\author{L. Clavelli}
\affiliation{Department of Physics and Astronomy, University of Alabama, Tuscaloosa, AL, USA}

\begin{abstract}
We present a model for the triggering of Supernovae Ia (SN Ia) by a phase transition to exact supersymmetry (susy) in the core of a white dwarf star.
The model, which accomodates the data on SN Ia and avoids the problems of the standard astrophysical accretion based picture, is based on string landscape ideas and assumes that the decay of the false broken susy vacuum is enhanced at high density. In a slowly expanding susy bubble, the conversion of 
pairs of fermions to pairs of degenerate scalars releases a significant amount of energy which induces fusion in the surrounding normal matter shell.
After cooling the absence of degeneracy pressure causes the susy bubble to collapse to a black hole of about 0.1 solar mass or to some
other stable susy object.
\end{abstract}

\maketitle

\thispagestyle{fancy}


%%%%%%%%%%%%%%%%%%%%%%%%%%%%%%%%%%
\section{Introduction} 
   This talk, based on recent work done with Peter Biermann at the University of Alabama, is organized according to the following outline. 
We first present the basic assumptions of the susy model.  In section II we discuss the standard astrophysical picture and the resulting puzzles.
In section III we show how the proposed susy model relates to string landscape and vacuum decay ideas.  In section IV we discuss the properties of the
susy picture as they relate to current data on SN Ia.  Section V is reserved for predictions of the model. 

We assume that we live in a local minimum of the string landscape consisting of a broken susy universe with partners of the standard model particles near the TeV scale.  We do not address the puzzle of why the apparent vacuum energy is so far below the susy breaking scale.
Our minimum is connected by an eventual tunneling event to an exact susy universe with the same number of degrees of freedom consisting of degenerate fermions and sfermions excited from a zero energy vacuum.  

We postulate that any broken susy system involving heavy nuclei (i.e. above helium) with a density distribution $\rho(\vec{x})$ has a probability per unit time to nucleate a critical exact susy bubble given by
\be
     \frac{dP}{dt} = \frac{1}{\tau_0 V_0} \int d^3x e^{-(\rho_c/\rho(\vec{x}))^3} \quad .
\label{assumption}
\ee 
The two free parameters of the model are a critical density, $\rho_c$, which we take to be in the range of white dwarf densities and a space time volume,
$\tau_0\,V_0$ which we take to be of order $0.1$ Gyr (Earth radius)$^3$.  As can be seen from eq.\,\ref{assumption} the transition rate grows rapidly with
$\rho(\vec{x})$ until it becomes of order $\rho_c$ after which the rate is proportional to the volume of the object at that or greater density.
Thus white dwarf stars are susceptible to the susy phase transition while much less dense objects such as the Earth and sun are not. Neutron stars
which have a much greater density but much less volume are also not susceptible.
Without loss of generality, $V_0$ can be chosen to be the maximum of the integral in eq.\,\ref{assumption} over all available systems in which case
$\tau_0$ becomes the minimum lifetime of such systems against the susy phase transition.
Sub critical bubbles which have a radius less than some critical radius, $R_c$ are immediately quenched but critical bubbles will grow as long as their
radius is greater than $R_c$.  As will be discussed in section III, the critical radius is density dependent in such a way that a bubble nucleated in a high density star will not grow beyond the boundary of the star.  It is much less probable for a critical bubble to be nucleated in the vacuum but, if that happens, it will grow without limit to convert the entire universe to a susy state.  In that sense, SN Ia are, in the current model, precursors to the
eventual vacuum decay of the universe.
    Since, in laboratory experiments, white dwarf densities are never produced over significant space time volumes, the assumption of eq.\,\ref{assumption}
about the behavior of matter at the high density frontier cannot be disproven in terrestrial experiments.  Its connection with string landscape and vacuum decay ideas is discussed in section III.

\section{Supernovae Ia and the standard astrophysical model}
  In the standard astrophysical model it is assumed than SN Ia can be explained by the theory of classical accretion onto white dwarf stars together with  standard nuclear fusion.  As a white dwarf star accretes matter from a companion star, it heats up until nuclear fusion is induced.  A natural product of
carbon and oxygen fusion is $^{56}$Ni which is clearly observed in the SN Ia spectra.  A large amount of this isotope must be produced before the accretion process reaches the Chandrasekhar limit of about 1.4 solar masses at which the star must collapse to a black hole.  Although thousands of white dwarf stars have been observed by the Sloan Digital Sky Survey (SDSS) including many in binary systems, none of these have the orbital characteristics to be a 
candidate for an accretion-induced supernova.  Since white dwarfs are predominantly carbon or oxygen, the prominence of $^{56}$Ni conclusively establishes white dwarfs as the source of SN Ia.  There are two potential routes to SN Ia in standard astrophysics, the single degenerate (SD) route whereby a white dwarf accretes
matter from a main sequence star such as the sun or a red giant and the double degenerate (DD) route whereby a white dwarf accretes from a second white dwarf.  Prior to 2010
the greatly preferred route was the SD route.  The obstacles standing in the way of the DD scenario were summarized in a 2000 review \cite{HillebrandtNiemeyer}:

 ``Besides the lack of convincing direct observational evidence for sufficiently many appropriate binary systems, the homogeneity of `typical' SNe Ia may be an argument against this class of progenitors. It is not easy to see how the merging of two white dwarfs of (likely) different mass, composition, and angular momentum with different impact parameters, etc, will always lead to the same burning conditions and, therefore, the production of a nearly equal amount of $^{56}$Ni.'' 

    However, accretion in the single degenerate scenario was predicted to produce significant amount of x-ray emission from the accreting matter.  A direct search for this emission yielded negative results and a comprehensive analysis \cite{GilfanovBogdan} concluded that

``no more than about five per cent of type Ia supernovae in early-type
galaxies can be produced by white dwarfs in accreting binary
systems, unless their progenitors are much younger than the bulk
of the stellar population in these galaxies, or explosions of sub-
Chandrasekhar white dwarfs make a significant contribution to
the supernova rate.''

A similar conclusion is reached by \cite{Bianco} due to the absence of expected effects from the early ejecta from the white dwarf striking a main sequence companion.

Theoretical work on the standard astrophysical models dates back decades and continues unabated but it is clear that
both types of accretion models have severe problems.  In addition, since no suitable binary systems have been seen and there is no data on how many might exist, the standard models are unable to make rate or timing predictions for SN Ia in sharp contradistinction to the model proposed here.  

The phase transition model we propose is an explosion of a sub-Chandrasekhar white dwarf independent of whether or not there is a binary partner.

\section{The String Landscape and Vacuum Decay Connection}

The transition to exact susy is an example of a decay of a false vacuum.

\begin{figure}[ht]
\includegraphics[width=80mm]{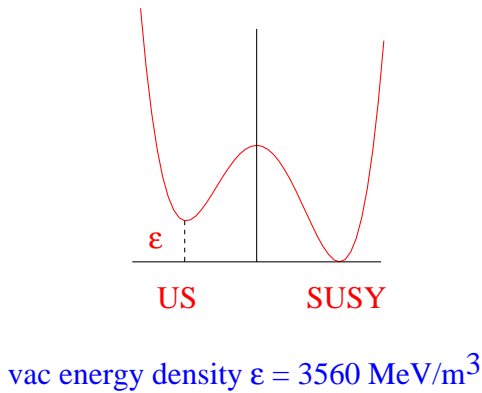}
%\mbox{\resizebox{288pt}{!}{\includegraphics{well.eps}}}
\caption{Double well potential with a metastable broken susy minimum and an exact susy true vacuum with zero vacuum energy.}
\label{well}
\end{figure}

In the vacuum the probability per unit time per unit volume for the decay of the broken susy vacuum is governed by the Coleman-DeLuccia formula \cite{Coleman}.
%(Coleman-De Luccia 1980)

\be
     \frac{d^2P}{dt d^3 r} \; = \; A_C e^{-B(vac)}  
\label{AemB}
\ee

\noindent with

\be
%\rho_c = \left( \frac{27 \pi^2 S^4}{2\,\hbar \, c} \right)^{1/3} \quad .
B(vac) \;  = \; \frac{27 \pi^2 S^4}{2\,\hbar\,c\,\epsilon^{3} } \quad .
\label{AemB2}
\ee

$A_C^{-1} \, = \tau_0 \, V_0$ is a parameter with the units of a space-time volume.  $S$ is the surface tension of the true vacuum bubble in the dominantly false vacuum background.  In the case of a susy phase transition $\epsilon$ is the difference between the vacuum energy density of our universe and the zero vacuum energy density of an exactly supersymmetric background.

There are many reasons to expect that the transition would be accelerated instead of impeded by matter effects  \cite{inducedvacdecay} .  These range from thermal effects that 
make it easier to tunnel through the barrier to density effects.  The latter have been proven \cite{Gorsky} to occur in lower dimensions but could also be expected to be effective in 4D.  One could also note that the tunneling probability of eq.\,\ref{AemB} is an increasing function of the vacuum energy difference between the two minima.  The tunneling depends on the total energy density difference at least in a model where the
scalar field defining the potential is coupled to the trace of the full energy momentum tensor.

\begin{figure}[ht] 
\includegraphics[width=80mm]{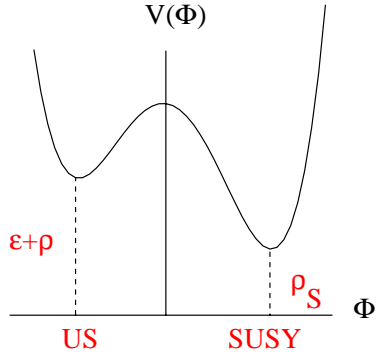}
\caption{The susy phase diagram in regions of dense matter}
\label{susywell}
\end{figure}

We therefore explore the assumption that the matter effect can be described by replacing the vacuum energy density dependence by the full energy density
difference between the two states.  
 
\be
      B(matter) = \frac{27 \pi^2 S^4}{2\,\hbar\,c\,(\epsilon+\Delta\rho\,c^2)^3 } \quad .
\label{Bmatter}
\ee

In exact susy, the degeneracy of bosons and fermions plus the availability of a conversion mechanism \cite{CK05,growth} 
%(Clavelli \& Karatheodoris 2005; Clavelli \& Perevalova 2005) 
from a pair of fermion to a pair of partner scalars implies that $\Delta \rho \, c^2$ is the Pauli excitation energy density of the fermions. The Fermi gas model predicts for this excitation energy in a heavy nucleus $\Delta \rho \equiv \rho - \rho_S \approx 0.02 \rho$.  This is about three times the energy release in hydrogen fusion and much more than that of carbon fusion which  fuels SN Ia in the standard picture.
 
In dense matter containing heavy nuclei (above helium in which all nucleons are in the lowest level), $\epsilon$ is negligible compared to $\Delta \rho\,c^2$.
and we would therefore arrive at the basic assumption given in eq.\,\ref{assumption}.

In the vacuum a critical bubble, one that will grow to take over the universe, is one whose radius is greater than a critical radius
\be
    R_c(vac) = \frac{3 S}{\epsilon} \quad .
\ee
In matter it is natural to assume that the critical radius is density dependent:
\be
    R_c(matter) = \frac{3 S}{\epsilon + \Delta \rho\,c^2} \quad .
\ee
A consequence is that a critical bubble nucleated in a homogeneous region will grow without limit while a critical bubble nucleated in dense matter
can become sub-critical and cease to grow if it reaches a boundary of the dense region.

\section{Consequences of a susy transition in a dense star}

In a white dwarf star of given composition, the density distribution is dependent only on the mass of the star according to the calculation first done by Chandrasekhar.  Thus, from eq.\,\ref{assumption}, each white dwarf star, has a mass-dependent lifetime $\tau$ assuming accretion is not a dominant effect.
 
\begin{figure}[ht] 
\includegraphics[width=80mm]{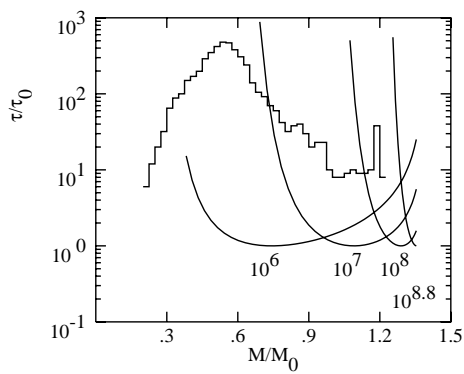}
\caption{The white dwarf lifetime relative to the free parameter $\tau_0$ as a function of mass relative to solar mass $M_\odot$ for several choices of the critical density parameter, $\rho_c$.  We superimpose the distribution of white dwarf masses from the Sloan Digital Sky survey.}
\label{tauofM}
\end{figure}

As can be seen from figure \ref{tauofM}, if the critical density is greater than about $10^7$ g/cc, only the high mass tail of the white dwarf spectrum
decays with a lifetime comparable to $\tau_0$ which we find in a rough fit equal to about $0.5$ Gyr.

Unlike the case of the standard astrophysical model where the distribution of suitable progenitors is unknown, we are able to fit the rate of SN Ia
using the known single white dwarf distribution data.  If we ignore the rate of new production of white dwarfs, this is

\be
    \frac{dN_{SN Ia}}{dt} = \int dM \frac{dN_{WD}}{dM} \frac{1}{\tau(M)} \quad .
\ee

\begin{figure}[ht] 
\includegraphics[width=80mm]{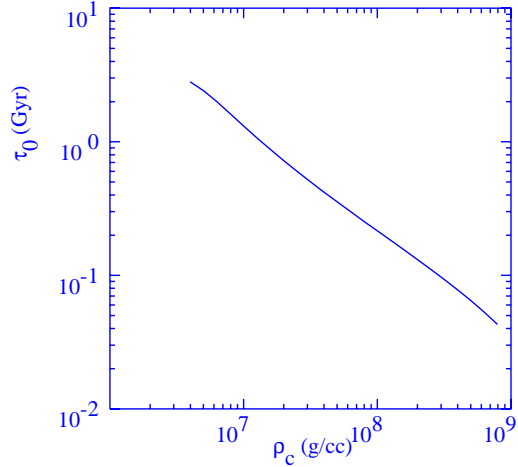}
\caption{The relation between the two free parameters of the model required to fit the average SN Ia rate.}
\label{rhogf}
\end{figure}

Fitting to the known rate of about 1 SN Ia per century in an average galaxy gives us the relation shown in figure\,\ref{rhogf} between the two free parameters of the model.  For a critical density of about $3 \cdot 10^{7}$ g/cc, $\tau_0 \approx 5\cdot 10^8$ yr.

However, the SN Ia rate is found to be a function of both the total mass of
stars in a galaxy and the rate of star production
\cite{ScannapiecoBildsten}:
\be
     \frac{dSNIa}{dt} = A \frac{M_{\mathrm gal}}{10^{10} M_\odot} + B \frac{\frac{dM_{\mathrm gal}}{dt}}{10^{10}M_\odot/{\mathrm Gyr}}
\label{Scannapieco}
\ee
with $A=0.04_{0.011}^{+0.016}$ and $B=2.6 \pm 1.1$.\\

The observation that the supernova rate is greatly enhanced in galaxies with elevated star production rates suggests that there is a component of the supernova rate at small delay times.  The observation that supernovae continue to be produced in older galaxies suggests that there
is also a component with longer delay times.  Standard model approaches to SNIa strive to be compatible with this observation leading perhaps to a bi-modal delay time hypothesis but no firm prediction.

In our model there is a 
natural spike in the lifetime distribution near some minimum $\tau_0$ 
plus a tail to much higher lifetimes. 

The rate of change of the number, $N$, of white dwarfs per unit mass at given time after galaxy formation is the difference between a production rate and a decay rate:
\be
    \frac{dN}{dt} = \frac{dN_{\displaystyle{prod}}}{dt} - \frac{N}{\tau(M)}\quad .
\ee
The supernova rate in this model per unit progenitor mass is
\be
     \frac{dN_{SN Ia}}{dt} = \frac{N}{\tau(M)} \quad .
\ee
Since the white dwarf production rate is expected to be roughly proportional to the overall star production rate the behavior in eq.\,\ref{Scannapieco} is
compatible with the current model.

As can be seen from figure \ref{tauofM}, There is a component of SN Ia with lifetimes close to the minimum in the $\tau$ vs $M$ plane.  In star-burst galaxies white dwarfs with the favored mass will rapidly decay creating a prompt component of the SN Ia distribution.  White dwarfs of other masses will decay with longer lifetimes. When
star production slows and white dwarfs at the favored mass have been depleted, dwarfs of other masses will produce supernovae yielding the effect observed in \cite{ScannapiecoBildsten}.  A prediction of the current model is that, in galaxies with enhanced star production, the supernovae will have a
narrow distribution in progenitor mass and ejected mass.  In older galaxies, there will be a broader distribution of progenitor masses.
As greater statistics of
high mass white dwarfs are accumulated, we would predict a dip in the white dwarf mass distribution at the minimum of the $\tau/\tau_0$ curve for chosen critical density.

     The current model, suggesting new physics effects when matter achieves white dwarf density, can also shed light on the puzzling gap in the black hole mass distribution.  Observations indicate that there are few, if any, black holes with masses between $10 M_\odot$ and $10^{5}\,M_\odot$.  Since the Schwarzschild radius is
\be
    R_S = 2\,G_N\,M/c^2 \,= \,4.64\,10^{-4}\,R_E\,M/M_\odot ,
\ee
the maximum density that any mass $M$ stellar conglomeration can achieve without becoming a black hole is 
\be
\rho_{max}=\frac{3\,M}{4\,\pi\,{R_S}^3}=\rho_{WD}\,\left(\frac{10^5\,M_\odot}{M}\right)^2
\ee
where, since the typical white dwarf has about solar mass and earth radius, we have defined a nominal white dwarf density 
\be
\rho_{WD}= \frac{3\,M_\odot}{4\pi\,R_E^3}\quad .
\ee
The suggestion, then, is that conglomerations of mass greater than $10^5\,M_\odot$ become black holes before achieving the critical density for a susy phase transition.  Conglomerations of smaller mass undergo the susy phase transition before becoming a black hole.  In the gap region, the phase transition causes the boiling off of most of the mass leaving a black hole of
much smaller mass.
\section{Consequences of a susy transition in a dense star}
The susy phase transition model with two free parameters provides a useful extra source of energy for SN Ia explosions.  It suggests the collapse of isolated white dwarfs which are absolutely stable in the standard astrophysical models. Predicted lifetimes range from some minimum $\tau_0$ to much larger values.  For large values of the critical density parameter
a narrow range of progenitor masses is predicted thus justifying the
use of SN Ia as standard candles for distance measurements.  In terms of the two parameters the SN Ia rate among other features can be fit.  Every SN Ia is
predicted to leave behind a small susy object (susy black hole?) with a mass equal to a small fraction of solar mass.  These objects can excite new radiation if they collide with other objects or heavy nuclei in the interstellar gas.
The model predicts an edge in the black hole mass distribution near $10^5\,M_\odot$.

    The model requires that there be two susy phases, exact susy and broken susy.  If the broken susy phase does not exist the model can be discarded. 
Otherwise, the entire universe is expected to eventually become supersymmetric due to the vacuum nucleation of a bubble of exact susy.  Since the probability per unit time for such a nucleation in volume, V, is proportional to V and the
volume of the universe is growing exponentially with time constant
\be
    \tau  = \frac{1}{\sqrt{24 \pi G_N \epsilon}}= 5.6\cdot 10^{9} yr 
\ee
the scale of the future lifetime of our broken susy universe (as well as the past lifetime) is expected to be of this order of magnitude.

%\begin{acknowledgments}
This research was supported in part by the DOE under grant DE-FG02-10ER41714.  
%\end{acknowledgments}
%\begin{thebibliography}{9}   % Use for  1-9  references

\end{document}